\documentclass[aps,prl,preprint,groupedaddress,showpacs,showkeys]{revtex4-1}
\usepackage{graphicx,amsmath,amssymb}

\newcommand{\degK}{^\circ\mathrm{K}~}
\newcommand{\mbo}[1]{$#1$}
\newcommand{\MPl}{M_{\rm Pl}}
\newcommand{\mpl}{M_{\rm Pl}}
\newcommand{\tpl}{t_{\rm Pl}}
\newcommand{\tCMB}{t_{\rm CMB}}
\newcommand{\ly}{~\ell \mathrm{y}}
\newcommand{\power}[1]{\times 10^{#1}}
\newcommand{\crn}{\nn \\ }
\newcommand{\bea}{\begin{eqnarray}}
\newcommand{\eea}{\end{eqnarray}}
\newcommand{\Ba}{\begin{eqnarray}}
\newcommand{\Ea}{\end{eqnarray}}

\newcommand{\nn}{\nonumber}

\newcommand{\eps}{\varepsilon}

\newcommand{\gv}{\mbox{GeV}}

\newcommand{\epo}{\,.}
\newcommand{\semis}{\,;\;\;}
\newcommand{\comas}{\,,\;\;}
\newcommand{\be}{\begin{equation}}
\newcommand{\ee}{\end{equation}}

\newcommand{\MSb}{$\overline{\rm MS}$ }

\newcommand{\D}{\mathrm{d}}

\begin{document}


\title{%
\vskip-3cm{\baselineskip14pt
\centerline{\normalsize DESY~13-093,~~HU-EP-13/25\hfill}
}
\vskip1.5cm
The hierarchy problem of the electroweak Standard Model revisited}


\author{Fred Jegerlehner}
\email[]{fjeger@physik.hu-berlin.de}
\homepage[]{www-com.physik.hu-berlin.de/\~{}fjeger/}
\affiliation{Humboldt-Universit\"at zu Berlin, Institut f\"ur Physik,
       Newtonstrasse 15, D-12489 Berlin, Germany\\
Deutsches Elektronen-Synchrotron (DESY), Platanenallee 6, D-15738 Zeuthen, Germany}


\date{\today}

\begin{abstract}
A careful renormalization group analysis of the electroweak Standard
Model, considered as a low energy effective theory, reveals that there
is no hierarchy problem in the broken phase of the SM. In the broken
phase a light Higgs turns out to be natural as it appears self-protected
and self-tuned by the spontaneous symmetry breaking. It means that the scalar Higgs
needs not be protected by any extra symmetry, specifically super
symmetry, in order not to be much heavier than the other SM particles,
which are protected by gauge- or chiral-symmetry.  Thus the existence
of quadratic cutoff effects in the SM cannot motivate the need for a
super symmetric extensions of the SM, but in contrast plays an
important role in triggering the electroweak phase transition at a
scale about $\mu_0\sim 10^{17}~\gv$ and in shaping the Higgs potential
in the early universe to drive inflation as supported by
observation. The impact on the inflation profile of the quadratically
enhanced bare Higgs mass term in the unbroken phase is discussed in
some detail. My analysis suggests that inflation in the early universe
is a direct consequence of the SM Higgs sector with its quadratic enhancement
of the bare Higgs mass term.
\end{abstract}


\maketitle



After the Higgs discovery by ATLAS~\cite{ATLAS} and CMS~\cite{CMS} at
the LHC essentially all ingredients of the Standard Model (SM) of
elementary particles are experimentally established and given the
Higgs mass $M_H=125.5\pm1.5~\gv$ for the first time all relevant SM
parameters are determined with remarkable accuracy~\cite{pdg}. It also
is quite commonly accepted that the SM is a low energy effective
theory of a system residing at the Planck scale and exhibiting the
Planck scale as a physical cutoff (see e.g.~\cite{Jegerlehner:1976xd}
and references therein on how to construct a renormalizable low energy
effective field theory from a cutoff theory). Strictly speaking we
consider the cutoff theory to be the fundamental theory, without
knowing what it is and how precisely the cutoff enters, we only know
its long range tail, the SM, for sure. One possible realization is a
lattice version of the SM, in the same spirit as lattice QCD. It is
then possible to predict effective bare parameters of the cutoff
system form SM properties. Extending arguments presented in
Ref.~\cite{Jegerlehner:2013cta}, we show that a consequence of the SM
Higgs mechanism is that there is no hierarchy or
naturalness~\cite{'tHooft:1979bh} problem in the SM concerning the
value of the Higgs mass. Small masses are
natural only if setting them to zero increases the symmetry of the
system. By applying the appropriate matching conditions to transfer
physical on shell parameters to corresponding \MSb ones (see
e.g.~\cite{Jegerlehner:2012kn} and references therein), together with
up-to-date \MSb renormalization group equations one can predict the
evolution of SM parameters up to the Planck scale, as a result of an
intricate conspiracy of SM parameters. In the following we assume all
parameters considered to be
\MSb parameters if not specified otherwise. As usual, by $\mu$ we
denote the \MSb renormalization scale. Actually, except for the
Abelian $U(1)_Y$ coupling $g'$ all other couplings turn out to behave
asymptotically free. The Higgs self-coupling $\lambda$ turns into an
asymptotically free parameter due to the large top Yukawa coupling
$y_t$, while the top Yukawa coupling is transmuted to be
asymptotically free by the large QCD coupling $g_3$. Thus both
$\lambda$ and $y_t$ are other-directed as part of the SM, such that no
strong coupling problems show up below the Planck scale.  In the
broken phase, characterized by the non-vanishing Higgs field vacuum
expectation value (VEV) $v(\mu^2)$, all the masses are determined by
the well known mass-coupling relations
\begin{eqnarray}
 m_W^2(\mu^2)&=&\frac14\,g^2(\mu^2)\,v^2(\mu^2)\semis
m_Z^2(\mu^2)=\frac14\,(g^2(\mu^2)+g'^2(\mu^2))\,v^2(\mu^2)\semis\crn
m_f^2(\mu^2)&=&\frac12\,y^2_f(\mu^2)\,v^2(\mu^2)\semis
m_H^2(\mu^2)=\frac13\,\lambda(\mu^2)\,v^2(\mu^2)\epo
\label{vevsquare}
\end{eqnarray}
The RG equation for $v^2(\mu^2)$ follows from the RG equations
for masses and massless coupling constants using one of these relations.
As a key relation we use~\cite{Jegerlehner:2001fb,Jegerlehner:2002er,Jegerlehner:2002em}
\begin{eqnarray}
\mu^2 \frac{d}{d \mu^2} v^2(\mu^2)
=3\, \mu^2 \frac{d}{d \mu^2} \left[\frac{m_H^2(\mu^2)}{\lambda(\mu^2)} \right]
\equiv
v^2(\mu^2) \left[\gamma_{m^2}  - \frac{\beta_\lambda}{\lambda} \right]\,,
\label{vev}
\end{eqnarray}
where $\gamma_{m^2} \equiv \mu^2 \frac{d}{d \mu^2} \ln m^2$ and
$\beta_\lambda \equiv \mu^2 \frac{d}{d \mu^2} \lambda \,.$ We write
the Higgs potential as $V=\frac{m^2}{2}\,H^2+\frac{\lambda}{24}H^4$,
which fixes our normalization of the Higgs self-coupling. When the
$m^2$-term changes sign and $\lambda$ stays positive, we know we have
a first order phase transition. The vacuum jumps from $v=0$ to $v\neq
0$. Such a phase transition happens in the early universe after the
latter has cooled down as a result of the expansion.

We remind that all dimensionless couplings satisfy the same RG
equations in the broken and in the unbroken phase and are not affected
by any quadratic cutoff dependencies. The evolution of SM couplings in
the \MSb scheme up to the Planck scale has been investigated in
Refs.~\cite{Hambye:1996wb,Holthausen:2011aa,Yukawa:3,degrassi,Moch12,Mihaila:2012fm,Chetyrkin:2012rz,Masina:2012tz,Bednyakov:2012rb,Bednyakov:2012en,Bednyakov:2013eba,Chetyrkin:2013wya,Tang:2013bz,Buttazzo:2013uya}
recently, and has been extended to include the Higgs VEV and the
masses in Refs.~\cite{Jegerlehner:2012kn,Jegerlehner:2013cta}. Except
for $g'$, which increases very moderately, all other couplings
decrease and stay positive up to the Planck scale. This 
strengthens the reliability of perturbative arguments and reveals a
stable Higgs potential up to the Planck
scale~\cite{Jegerlehner:2012kn,Jegerlehner:2013cta}. While most
analyses~\cite{Yukawa:3,degrassi,Moch12,Masina:2012tz,Buttazzo:2013uya}
are predicting that for the found Higgs mass value vacuum stability is
nearby only (metastability), and actually fails to persist up to the Planck scale, our
evaluation of the matching conditions yields initial \MSb parameters
at the top quark mass scale which evolve preserving the positivity of
$\lambda$. Thereby the critical parameter is the top quark Yukawa
coupling, for which we find a slightly lower value. In view of the
fact that the precise meaning of the experimentally extracted value of
the top quark mass is not free of ambiguities, usually it is
identified with the on-shell mass $M_t$ (see
e.g.~\cite{Jegerlehner:2012kn} and references therein), it may be
premature to claim that instability of the SM Higgs potential is a
proven fact already. I also think that the implementation of the
matching conditions is not free of ambiguities, while the evolution of
the couplings over many orders of magnitude is rather sensitive to the
precise values of the initial couplings. Accordingly, all numbers
presented in this article depend on the specific input parameters
adopted, as specified in Ref.~\cite{Jegerlehner:2012kn,Jegerlehner:2013cta}. In
case the Higgs self-coupling has a zero $\lambda(\mu^2)=0$ at some
critical scale $\mu_c$ below $\MPl$ we learn from Eq.~(\ref{vev}), or
more directly from $v(\mu^2)=\sqrt{6m^2(\mu^2)/\lambda(\mu^2)}\stackrel{\lambda \to +0}{\to}
\infty$ that the SM looses it meaning above this singular point.

The most serious problem in the low energy effective SM is the
hierarchy problem caused by the quadratic divergences in the Higgs
mass parameter, which are the same in the symmetric as well as in the
broken phase, since spontaneous breaking of the symmetry does not
affect the ultraviolet (UV) structure of the theory. This does not
rule out that the effective bare quantities depend on screening
effects of the couplings in the low energy effective theory as we will
see. Quadratic divergences have been investigated at one loop in
Ref.~\cite{Veltman:1980mj} (see also~\cite{Decker:1979cw}), at two
loops in Refs.~\cite{Alsarhi:1991ji,Hamada:2012bp}. Including up to
$n$ loops the quadratic cutoff dependence, which in dimensional
regularization (DR) shows up as a pole at $D=2$, is known to be given
by
\be
\delta m_H^2= \frac{\Lambda^2}{16\pi^2}\,C_n(\mu)
\label{quadraic1}
\ee
where the n-loop coefficient only depends on the gauge couplings $g'$,
$g$, $g_3$, the Yukawa couplings $y_f$ and the Higgs self-coupling
$\lambda$. Neglecting the numerically insignificant light fermion
contributions, the one-loop coefficient function $C_1$ may be written
as
\bea
C_1=2\,\lambda+\frac32\, {g'}^{2}+\frac92\,g^2-12\,y_t^2
\label{coefC1}
\eea
and is uniquely determined by dimensionless couplings. The latter are
not affected by quadratic divergences such that standard RG equations
apply. Surprisingly, as first pointed out in
Ref.~\cite{Hamada:2012bp}, taking into account the running of the SM
couplings, the coefficient of the quadratic divergences of the bare
Higgs mass correction can vanish at some scale. In our analysis we get
a scenario where $\lambda(\mu^2)$ stays positive up to the Plank scale
and looking at the relation between the bare and the renormalized
Higgs mass we find $C_1$ and hence the Higgs mass counterterm to
vanish at about $\mu_0\sim 7 \times 10^{16}~\gv$, not very far below
the Planck scale. It also has been shown in~\cite{Hamada:2012bp} that
the next-order correction
\bea
C_2&=&C_1+ \frac{\ln (2^6/3^3)}{16\pi^2}\, [
18\,y_t^4+y_t^2\,(-\frac{7}{6}\,{g'}^2+\frac{9}{2}\,g^2
             -32\,g_s^2) \nn \\ &&+\frac{77}{8}\,{g'}^4+\frac{243}{8}\,g^4
             +\lambda\,(-6\,y_t^2+{g'}^2+3\,g^2)
             -\frac{10}{18}\,\lambda^2]
\label{coefC2}
\eea
does not change significantly the one-loop result. The same results apply
for the Higgs potential parameter $m^2$, which corresponds to
$m^2\hat{=}\frac12\,m_H^2$ in the broken phase. For scales $\mu <
\mu_0$ we have $\delta m^2$ large negative, which is triggering
spontaneous symmetry breaking by a negative bare mass $m_{\rm
bare}^2=m^2+\delta m^2$, where $m=m_{\rm ren}$ denotes a renormalized
mass. At $\mu=\mu_0$ we have $\delta m^2=0$ and the sign of $\delta
m^2$ flips, implying a phase transition to the symmetric phase. Finite
temperature effects, which should be included here in a realistic
scenario, will be addressed briefly at the end of an Addendum, where
we argue that finite temperature effects do not change the gross
features of our scenario. 

Such a phase transition is particularly relevant for inflation
scenarios in the evolution of the early universe.  Going back in
cosmic time, at $\mu_0$ the Higgs VEV jumps to zero and SM gauge boson
and fermion masses all vanish, at least provided the scalar
self-coupling $\lambda$ continues to be positive as inferred in recent
analyses, like in~\cite{Jegerlehner:2013cta}. Note that the phase
transition scale $\mu_0$ is close to the zero $\mu_\lambda\sim
3.5\power{17}$ of $\beta_\lambda$,
i.e. $\beta_\lambda(\mu_\lambda)=0$. While $\lambda$
is decreasing below $\mu_\lambda$ it starts to increase weakly above
that scale.

Now considering the hierarchy problem: it is true that in the relation
\be
m^2_{H\,{\rm bare}}=m^2_{H\,{\rm ren}}+\delta m_H^2
\label{tuning}
\ee 
both $m^2_{H\,{\rm bare}}$ and $\delta m^2_{H}$ are many many orders
of magnitude larger than $ m^2_{H\,{\rm ren}}\,.$ Apparently a severe
fine tuning problem. However, in the broken phase $ m^2_{H\,{\rm
ren}}\propto v^2(\mu^2_0)$ is $O(v^2)$ not $O(\MPl^2)$, i.e. in the
broken phase the Higgs is naturally light, as $v_{\rm
bare}=v(\mu_0^2)$ at scale $\mu=\mu_0$. That the Higgs mass likely is
$O(\MPl)$ in the symmetric phase is what realistic inflation scenarios
favor (see below). The light Higgs self-protection mechanism in the
broken phase only works because of the existence of the zero of the
coefficient function in front of the huge (but finite) prefactor in
Eq.~(\ref{quadraic1}), which defines a matching point at which
$m^2_{H\,{\rm bare}}=m^2_{H\,{\rm ren}}$. The renormalized value
$m^2(\mu^2)$ at $\mu_0$ evolves according to the renormalized RG
evolution equation and in fact remains almost constant between the
scale $\mu_0$ and $M_Z$ as has been shown in
Ref.~\cite{Jegerlehner:2013cta}.

Note that away from the phase transition point there is still a huge
cancellation in Eq.~(\ref{tuning}), however this cancellation is tuned
by the SM itself. Ihis can be understood to work in a similar way as
the non-Abelian gauge cancellations, which is known to be the result
of the low energy expansion where non-renormalizable terms are
suppressed by powers of the cutoff. Such self-organization is typical
in critical phenomena (see below). Looking at Eq.~(\ref{vevsquare}),
the key point is that in the broken phase all SM masses,
\textbf{including the Higgs mass}, are of the type $\propto g_i(\mu^2)
\times v(\mu^2)$ ($g_i=g,\sqrt{g^2+g^{'2}},y_t,\sqrt{\lambda}$) and it
is natural to have the Higgs in the ballpark of the other, so called
protected masses (gauge bosons by gauge symmetry and fermions by
chiral symmetry), since for all masses the scale is set by $v(\mu^2)$.
In fact, it is the Higgs system itself which sets the scale for all
masses. One should note that $v$ is an order parameter, like the
magnetization in a ferromagnetic Ising system consisting of up and
down spins on a lattice of spacing $a$, where neighboring parallel
spins attract each other while anti-parallel ones repel each other
with equal strength. The order parameter is not to be expected to be
of the order of the inverse lattice spacing but is a matter of the
strength of the nearest neighbor interactions and the collective long
range order and domain wall structure which emerges as a long range
phenomenon. In the SM, likewise, there is no reason to expect
$v=O(\mpl)$. Our interpretation, that the SM has no hierarchy problem
in the Higgs phase is in accord with the naturalness argument, namely
the symmetry get enhanced when we send $v$ to zero. This of course
does not say anything about issues like the unknown origin of the
hierarchy of the Yukawa couplings. Note that for $M_H=125~\gv$ at the
Planck scale $\mu=\MPl$ the relevant couplings have values $g'\simeq
0.46,\,g\simeq 0.51,\, g_3\simeq 0.49,\,y_t\simeq
0.36,\,\sqrt{\lambda}\simeq 0.41$ of very similar magnitude, i.e. they
appear ``quasi unified'', what looks natural as they emerge form the
one cutoff system. Note that it is $\sqrt{\lambda}$ which compares to
the other SM couplings, not $\lambda$, as can be seen by inspecting
the mass coupling relations. A quadratically enhanced Higgs mass could
only be obtained if the dimensionless Higgs self-coupling would be
itself proportional to $\Lambda^2$ which really looks quite
nonsensical. Perturbation theory at least suggests that $\lambda$ can
be affected by logarithms of $\Lambda$ only, which typically sum to
some moderate anomalous dimension. By the way, the fine tuning would
have to apply for all masses in the same way as the quadratic
divergences stick in the relation between $v^2_{\rm bare}$ and the
\MSb variant $v^2(\mu^2)$. A corresponding observation/conclusion has
been reached recently within the Abelian Higgs model in
Ref.~\cite{Malinsky:2012tp}.

We note that the Higgs mass is self-protected from being huge by the
fact that in the broken phase the Higgs mass is generated via the
Higgs condensate as any other particle getting its mass by the Higgs
mechanism. As it should be, such self-protection does not apply for
singlet Majorana neutrino mass terms. Large singlet neutrino masses
are expected to be responsible for mediating a sea-saw mechanism,
which is able to explain the smallness of the neutrino masses.

In our view the SM hierarchy problem is a ``problem'' of the symmetric
phase only, where the quadratic enhancement of the bare Higgs mass
term actually provides the profile of slow-roll 
inflation~\cite{Starobinsky:1980te,Guth:1980zm,Linde:1981mu}.
In addition the phase transition clearly stops the inflation phase.
The Higgs condensation could be responsible for the reheating phase.

A similar self-protection mechanism we may expect to work for the
vacuum energy, which, if generated by quantum corrections, has the
form
\be
\delta \rho_{\rm vac}= \frac{\Lambda^4}{(16\pi^2)^2}\,X_n(\mu)
\label{quartic}
\ee
with a dimensionless coefficient $X_n$, which depends on the number
$n$ of loops included. Again, in leading order (assuming $m^2
\ll \Lambda^2$), we expect it only to depend on the dimensionless SM
couplings controlled by the standard RG equations. Excluding a
classical background density the leading contribution is of two loop
order. In case
$$X_2(g'(\mu^2),g(\mu^2),g_3(\mu^2),y_t(\mu^2),\lambda((\mu^2))$$ has
a zero we would have a matching point where a $\Lambda^4$ contribution
is absent. It again would mean that at some value of $\mu$ the
quartically cutoff dependent term would vanish: $\delta \rho_{\rm
vac}=0$ and the bare and the renormalized vacuum density would agree
up to subleading terms, such that below that point one could get a
severely tamed vacuum density. A detailed analysis of the coefficient
function $X_2(\mu)$ is missing yet, but certainly would shed new light
on the cosmological constant problem, the most severe remaining fine
tuning problem of the SM. Also it may be important that there exists
contributions to the cosmological constant of either sign. It is
worthwhile to mention here that in condensed matter systems the ground
state energy is not in general determined by $\Lambda^4$ with a $O(1)$
coefficient (see e.g.~\cite{Volovik:2005zu} for a discussion in our
context).

A general remark is in order here: the SM hierarchy problem may be
understood/interpreted in different ways.  Frequently, the hierarchy
problem is formulated in a strong form, saying that if the Higgs mass
is not protected by a symmetry and hence we expect $M_H=O(\Lambda)$
and supposing all couplings are $O(1)$ then $M_H=O(\Lambda)$ implies
$M_i=O(\Lambda)$ $(i=H,W,Z,t,\cdots)$ for all masses. In other words,
why is the electroweak scale $v$ not just $\Lambda$? However, this
type of argument is rather formal. According to this kind of
interpretation, the term ``spontaneous symmetry breaking'' would
become quite meaningless if the breaking would be naturally at the
``hard'' scale and not at a much lower ``soft'' one, as it is
anticipated usually. It would mean that the symmetric phase is not
recovered at the hard scale. In effective theories one has to
distinguish between short range (e.g. lattice spacing $a\sim
\Lambda^{-1}$) order and long range order quantities, the latter
emerging from collective behavior as encountered in phase
transitions. The fact that criticality requires the temperature to be
tuned to its critical value does not mean that the critical
temperature is $T_c=O(1/a)$. Note that, in field theory language, the
reduced temperature $(T-T_c)/T_c$ is proportional to the renormalized
mass square $m^2_{\rm ren}=m^2_{\rm bare}-m^2_{c\,{\rm bare}}$, where
$m^2_{c\,{\rm bare}}$ is the critical bare mass for which the
renormalized mass is zero.  The key point is that a limit $\Lambda \to
\infty$ need not exist as $\Lambda$ is a given physical quantity. The
critical ``fine tuning'' $T\sim T_c$ is not a fine tuning problem
giving an answer to why $T_c \ll 1/a$. In typical cutoff systems
encountered in condensed matter physics an order parameter associated
with a first order phase transition, like the Higgs VEV $v$ in our
case, is by no means $O(\Lambda)$. Rather it is a matter of a
collective phenomenon of the system with infinitely many degrees of
freedom. Below the critical temperature, on the bare level, depending
on the given effective short range interaction between the intrinsic
degrees of freedom, the system is building up long range order and
domain structures. The critical temperature $T_c$ as well as an order
parameter like the magnetization $M$ are macroscopic quantities. Long
range effective quantities emerging in critical phenomena, are effects
we see when looking at a system from far away and do not simply
reflect the microscopic structure.  The emergence of long range
collective patterns is what I called self-tuning or self-protection
above. It is the natural case in critical or quasi-critical condensed
matter systems. So it is natural to have $v \ll \Lambda$ and unnatural
to expect $v\sim\Lambda$. In the SM, in addition, stetting $v=0$
enhances the symmetry in any case (the gauge- and chiral-one), in
spite the Higgs mass square persists getting corrections
$O(\Lambda^2)$, which in the symmetric phase boosts up the physical
$m_H$ to an $O(\Lambda)$ quantity.

Another typical example is lattice QCD, where all dimensionful
quantities per se are obtained in units of the lattice spacing
$a=\Lambda^{-1}$, which however does not mean that all quantities are
of order $\Lambda$. Considering typically, dimension one physical
lattice quantities $O_i=c_i/a$, like masses or a Higgs VEV, we do not
expect all $c_i =O(1)$. Actually in collective phenomena near second
order phase transition points, where we expect an effective continuum
field theory to parametrize the scene, we may expect a wide hierarchy
of results. We also have to keep in mind, that fixing the physical
parameters always requires renormalization at some point.  After all
we have no direct experimental access to the bare parameters of the
Planck system, and have to be satisfied being able to determine the
low energy effective parameters. Concerning the lattice setup of the
SM, the proper definition of the Higgs VEV $v$ may be obscured due to gauge
ambiguities. In order to avoid ghosts one should stick to the unitary
gauge and define $v$ as the order parameter of the discrete $Z_2$
symmetry $H\leftrightarrow -H$ of the physical Higgs potential.

We conclude that in contrast to common wisdom the quadratic cutoff
dependencies of SM renormalization plays a crucial role in triggering
the EW phase transition and in shaping the inflation potential
(see~\cite{Jegerlehner:2013cta}). The SM Higgs necessarily triggers
inflation due to the quadratic short distance enhancement when
near or at the Planck scale. This results because of the unavoidable
contribution of the Higgs to energy density and pressure:
\Ba
\rho=\frac12\,\left(\partial\phi\right)^2+V\comas p=\frac12\,\left(\partial\phi\right)^2-V\;.
\Ea
Since \mbo{\nabla \phi=0} on cosmological scales, the time
derivative $\dot{\phi}\equiv \D \phi/\D t$ is relevant only, thus \mbo{\partial\phi \to
\dot{\phi}}. If in the symmetric phase the mass term of the Higgs potential
$V(\phi)=\frac{m^2}{2}\phi^2+\frac{\lambda}{24}\phi^4$ is positive and
dominates we have \mbo{\dot{\phi}^2\ll V} such that $p\simeq-\rho$.
Thus, while scalar field models in general have time varying equation
of state $w=p/\rho$ with $w\geq -1$, before the phase transition at
$\mu_0$, the SM predicts $w\approx -1$ as observationally supported by
the recent Planck mission result
$w=-1.13^{+0.13}_{-0.10}$~\cite{Ade:2013zuv}. The equation of state
$w=p/\rho=-1$ corresponds to the cosmological constant, which in
addition must be large during inflation, in order to dominate the
radiation as well as other terms contributing to $\rho$ in the
Friedman equation
\Ba
H^2+\frac{k}{a^2}=\frac{8\pi}{3\mpl^2}\,\rho\epo
\Ea
Here, $a(t)$ is the time dependent scale function of the Friedman-Robertson-Walker metric 
$\D s^2=\left(c\,\D t\right)^2-a^2(t)\,
\left\{ \frac{\D r^2}{1-kr^2}+r^2\,\left(\D \theta^2+\sin^2 \theta\,\D
\varphi^2\right)\right\}$
describing the expansion of the universe, $k=0,1,-1$ for flat,
spherically closed and hyperbolic open geometry, and $H\equiv \dot{a}/a$ is
the Hubble ``constant''. Before the phase transition takes place we then have
\mbo{H^2\simeq \frac{4\pi m^2}{3\mpl^2}\,\phi^2}.

The need for inflation one observes most directly if looking at the
Cosmic Microwave Background (CMB) patterns. 
A serious issue here is the horizon problem. According to standard Big
Bang cosmology we see a much larger part of sky, all within the Hubble
horizon \mbo{d=d_0\sim t_0 \sim \frac{1}{H_0}\sim 4 \cdot 10^{10}}
light years ($\ell \mathrm{y}$), as a uniform distribution, than the
causal patch $D_\mathrm{CMB}\simeq 4 \cdot 10^5\ly$ at the time
\mbo{t_{\rm CMB} \simeq 400.000~\mathrm{yrs}} of decoupling, when the 
present Hubble horizon was of size \mbo{d_{\tCMB}\simeq 4\cdot
10^{7}\ly}. This causality problem is solved by inflation: if we can
achieve to have \mbo{\ddot{a}>0} such that \mbo{a} grows faster than
\mbo{t}, today's CMB sky was smaller than the causal patch at some
early time. Inflation also solves the flatness problem. What is
required to accomplish this is a large cosmological constant for
a limited epoch of time. Now we know that the SM together with its
specific parameters just provides automatically the inflation of the
early universe, which gets stopped at the phase transition point at
scale $\mu_0$ (for details see the Addendum).

In view of their role for inflation in the early universe, it is more
likely that the absence of quadratic divergences as tailored by super
symmetry would be more a problem than a solution of an existing
problem. Very different recent reasoning about naturalness the reader may find
in Refs.~\cite{Aoki:2012xs,Blanke:2013uia,Tavares:2013dga,Masina:2013wja}.

Addendum: some consistency checks of the SM inflation scenario (see~\cite{Linde:1990xn}). 
So far we have said nothing about the size of the Higgs field and whether it
is adequate to just check the parameters in the potential to decide
about the relative importance of the different terms. Therefore some
more details on the impact of the quadratic enhancement on the
inflation profile are in order here. As already mentioned, in
Ref.~\cite{Jegerlehner:2013cta} we have shown that for a Higgs mass of
about 125 GeV there is a phase transition at a scale about \mbo{\mu_0
\sim 7\power{16}~\gv} and at temperatures above this scale the SM is
in the symmetric phase in which the Higgs potential exhibits a huge
bare mass term of size
\bea
m^2\sim \delta m_H^2 \simeq \frac{\mpl^2}{16\pi^2}\,C(\mu=\mpl) \simeq \left(0.0356\,\mpl \right)^2\semis
m^2(\mpl)/\mpl^2\approx 1.23\power{-3}\epo
\eea
For large slowly varying fields, the field equation of motion
$\ddot{\phi}+3\,H \dot{\phi}=-\frac{\partial V}{\partial \phi}\equiv -V'$
simplifies to the slow-roll equation $3\,H\dot{\phi}=-V'$, which describes
a decay of the field. 

One expects that \mbo{V(\phi)} does not exceed a possible vacuum energy of size
\mbo{\mpl^4}. The initial value for \mbo{\phi_0=\phi(\tpl)} at
Planck time \mbo{\tpl} then should be bound by
\mbo{\phi_0=\mpl^2/m(\mpl)}. In general, if the initial value of \mbo{\phi} is
exceeding about \mbo{\frac15 \,\mpl} one can neglect \mbo{\ddot{\phi}}
in the field equation as well as the kinetic term
\mbo{\frac12\,\dot{\phi}^2} in the Friedman equation. We expect
inflation to start at Planck time
\mbo{t_i\equiv t_\mathrm{initial}=\tpl= 5.4 \power{-44}~\mathrm{sec}} and to stop at the phase transition
at \mbo{t_e\equiv t_{\rm end}=t_{\rm Higgs}\approx 4.7
\power{-41}~\mathrm{sec}}. In this high
damping regime \mbo{\phi(t)=\phi_0-\frac{m\,\mpl}{2\sqrt{3\pi}}\,t}
with \mbo{\phi_0\approx 3.43\power{20}~\gv} and \mbo{\phi(t_e)\approx 2.81\power{20}~\gv} and \mbo{a(t)=a_0\,\exp
\left(\frac{2\pi}{\mpl^2}\left[\phi_0^2-\phi(t)^2\right]\right)} with
\mbo{a_e/a_i\approx \exp (1636.2) \simeq 10^{710.59}\simeq
10^{10^{2.85}}}.

We may cross check this by looking at the leading behavior given by
\mbo{a(t)=\exp{\,\left(H(\phi)\,t\right)}}  with Hubble
constant \mbo{H(\phi)\approx \sqrt{\frac{4\pi}{3}}\,\frac{m\,\phi}{\mpl}} and
for \mbo{m/\mpl=0.0356} and \mbo{\phi=\mpl^2/m} we have
\mbo{H(\phi)\approx 3.73\power{43}} and \mbo{Ht_e\approx 1768.6} corresponding 
to an expansion factor \mbo{\exp (Ht)=10^{10^{2.89}}} well consistent
with the previous estimate. Note that as required by the horizon
problem in particular the exponent \mbo{Ht} is much larger than unity
if \mbo{\phi} exceeds the Planck mass at these times. Needed is
\mbo{N\simeq H t > 60} to solve the horizon
problem. The SM blow up exponent is given by
\Ba
N&=&\ln \frac{a(t_{\rm end})}{a(t_{\rm
initial})}=\int\limits_{t_i}^{t_e}\,H(t)\,\D t
=\int\limits_{\phi_i}^{\phi_e}\,\frac{H}{\dot{\phi}}\,\D \phi
=\frac{8\pi}{\mpl^2}\,\int\limits_{\phi_e}^{\phi_i}\,\frac{V}{V'}\,\D
\phi =H\,(t_e-t_i)\;,
\Ea
and \mbo{N=H\,(t_e-t_i)} is exact if \mbo{H=\mathrm{constant}}
i.e. when \mbo{\rho=\rho_\Lambda} is dominated by the cosmological
constant. In the symmetric phase \mbo{V/V'>0} and hence \mbo{\phi_i > \phi_e}.
Note that  rescaling of potential does not affect inflation, but the
relative weight of the terms is crucial.

The slow-roll criteria are usually tested by the coefficients
$\eps\equiv\frac{\mpl^2}{8\pi}\,\frac12\,\left(\frac{V'}{V}\right)^2$
and $\eta \equiv \frac{\mpl^2}{8\pi}\,\frac{V^{''}}{V}$: $\eps \ll 1$
ensures $p \simeq -\rho $, while \mbo{\eps,\eta \ll 1} ensure
slow-roll for a long enough time, maintaining
\mbo{\ddot{\phi}\ll3\,H\dot{\phi}}.  When slow-roll ends, \mbo{\phi}
oscillates rapidly about \mbo{\phi=0} and the oscillations lead to
abundant particle production reheating the universe.  For the SM Higgs
potential in the symmetric phase, denoting \mbo{z\equiv
\frac{\lambda}{6\,m^2}}, we have
\mbo{\frac{V}{V'}=\frac{\phi}{4}\,\left(1+\frac{1}{1+z\phi^2}\right)}
and thus
\mbo{\int\limits_{\phi_i}^{\phi_e}\,\frac{V}{V'}\,\D \phi= \frac18\,\left[\phi_e^2-\phi_i^2
+\frac{1}{z}\,\ln \frac{
 {\phi_e}^2\,z+1}{{\phi_i}^2\,z+1}\right]}
or
\mbo{N=-\frac{\pi}{\mpl^2}\,\left[\phi_e^2-\phi_i^2-\frac{1}{z}\,\ln \frac{
 {\phi_e}^2\,z+1}{{\phi_i}^2\,z+1}\right]} 
which implies \mbo{\phi_i^2-\phi_e^2+\frac{1}{z}\,\ln \frac{
 {\phi_i}^2\,z+1}{{\phi_e}^2\,z+1}=\mpl^2\cdot N/\pi}.
Inflation requires \mbo{\phi_i > \phi_e} and the CMB horizon issue
\mbo{N} large, i.e., \mbo{\phi_i \gg \phi_e} and one can solve the
relation for $\phi_i$ starting with $\phi_e\sim 0$ and $N$ its
Gaussian approximation $N\sim 1636.2$ estimated before. While in the Gaussian
approximation $\phi_i=2.786\power{20}~\gv$ including the effective
Higgs coupling yields $\phi_i\approx 2.773\power{20}~\gv$, i.e. as
expected the solution is pretty stable, meaning that $V(\phi)\simeq
\frac{m^2}{2}\,\phi^2$ is truly what counts during the epoch of inflation.
With this solution as an initial value then evaluated at \mbo{t_e}
yields \mbo{\eps\approx 6\power{-4}} and
\mbo{\eta\approx9\power{-4}}, when reheating is triggered by the phase
transition near $\mu_0$. This can be used to
estimate the scalar density fluctuations:
\mbo{\delta \rho=\frac{\D V}{\D \phi}\,\delta\phi} exhibiting a spectrum
\mbo{A^2_s(k)=\left.\frac{V^3}{\mpl^6\,(V')^2}\right|_{k=aH}}
to be evaluated at the moment when the physical scale of the
perturbation \mbo{\lambda=a/k} is equal to the Hubble radius \mbo{H^{-1}}.
Observations are parametrized by a power spectrum
\mbo{A^2_s(k)\propto k^{n_s-1}} where
\mbo{n_s=1-6\eps+2\eta}. With the above rough estimates we 
find \mbo{n_s\approx 0.998} which confronts with the recent Planck
mission result \mbo{n_s=0.9603\pm0.0073}~\cite{Ade:2013uln}. I have not yet estimated the
uncertainty, which however is expected to be large enough not to be in
plain conflict with the data.

For what concerns the electroweak phase transition, it is important to
look at finite temperature
effects~\cite{FiniteTemp1,FiniteTemp2,FiniteTemp3,Dine:1992wr}. The leading modification
caused by finite temperature effects enters the finite temperature
effective potential $V(\phi,T)$: while at zero temperature
\mbo{V(\phi,T=0) = -\frac{\mu^2}{2}\,\phi^2+\frac{\lambda}{24} \,\phi^4}
at finite temperature we have
\mbo{V(\phi,T) =
\frac{1}{2}\,\left(g_T\,T^2-\mu^2\right)\,\phi^2+\frac{\lambda}{24}
\,\phi^4}. Usually it is assumed that the Higgs is in the broken phase
($\mu^2>0$) and that the EW phase transition is taking place when the
universe is cooling down below the critical temperature
$T_c=\sqrt{\mu^2/g_T}$. However, above the scale $\mu_0$ we are in the
symmetric phase with \mbo{-\mu^2\to m^2=m_H^2+\delta m_H^2}. As
claimed before, the phase
transition is triggered by \mbo{\delta m_H^2} with
\mbo{m^2\simeq1.27\power{-3}\,\mpl^2}.  In our case we obtain
\mbo{T(\mu=\mu_0)\simeq 8.12\power{29}~\degK} and
\mbo{T(\mu=\mpl)\simeq 5.04\power{30}~\degK} such that we expect
the EW phase transition to be triggered by the bare Higgs mass in
spite of the fact that the finite temperature term $g_T\,T^2$ is very
large in the early universe. In the SM $g_T$ is represented by SM
effective couplings.  In fact
\mbo{g_T=\frac{1}{4v^2}\,\left(2m_W^2+m_Z^2+2m_t^2+\frac12\,m_H^2\right)
=\frac{1}{16}\,\left[3\,g^2+{g'}^{2}+4\,y_t^2+\frac23\,\lambda
\right]} and taking the effective couplings at $\mpl$: 
$g'= 0.4561,\, g =0.5084,\, g_3= 0.4919 \pm 0.0046,\, y_t = 0.3551 \pm
0.0037,\,$ and $\lambda $ in the range $0.0896 \div 0.1648$. We then
estimate $g_T \approx 0.0983\sim 0.1$. Therefore, away from the phase
transition point, where $m^2_{\rm bare}=m^2_{\rm ren}(\mu^2_0)$, the
bare mass term is dominating anyway. Of course, our rough estimates
are no substitute for a more careful reanalysis of the EW phase
transition.

\bigskip

\noindent
\textbf{Acknowledgments:}\\
I am grateful to Oliver B\"ar, Mikhail Kalmykov and Daniel Wyler for
inspiring discussions.


\end{document}